\documentclass[prb,twocolumn,nopacs,superscriptaddress]{revtex4-2}
\usepackage[utf8]{inputenc}
\usepackage[american,]{babel}
\usepackage[T1]{fontenc}
\usepackage[pdftex]{graphicx}  
\usepackage{graphicx, xcolor}
\usepackage{dcolumn}
\usepackage{bm}
\usepackage{dsfont}
\usepackage{amsmath,amsthm,amssymb}
\usepackage{hyperref}
\usepackage[T1,T2A]{fontenc}
\usepackage{xcolor}
\hypersetup{colorlinks,bookmarksopen,bookmarksnumbered,
    citecolor=blue,
    linkcolor=blue,
    pdfstartview=false,
    urlcolor=blue}
\usepackage{graphicx}
\usepackage{braket}
\usepackage{wrapfig}
\usepackage{soul}
\usepackage{mathtools}
\usepackage{float}
\usepackage{tabularx}
\usepackage{physics}

\begin{document}

\title{Density and current statistics in boundary-driven monitored fermionic chains} 

\author{Xhek Turkeshi}
\affiliation{JEIP, UAR 3573 CNRS, Coll\`{e}ge de France, PSL Research University, 11 Place Marcelin Berthelot, 75321 Paris, France}
\author{Lorenzo Piroli}
\affiliation{Dipartimento di Fisica e Astronomia, Universit\`a di Bologna, via Irnerio 46, I-40126 Bologna, Italy}
\author{Marco Schir\`o}
\affiliation{JEIP, UAR 3573 CNRS, Coll\`{e}ge de France, PSL Research University, 11 Place Marcelin Berthelot, 75321 Paris, France}
\date{\today}

\begin{abstract}
We consider a one-dimensional system of non-interacting fermions featuring both boundary driving and continuous monitoring of the bulk particle density. Due to the measurements, the expectation values of the local density and current operators are random variables whose average behavior is described by a well studied Lindblad master equation. By means of exact numerical computations, we go beyond the averaged dynamics and study their full probability distribution functions, focusing on the late-time stationary regime. We find that, contrary to the averaged values, the spatial profiles of the median density and current are non-trivial, exhibiting qualitative differences as a function of the monitoring strength. At weak monitoring, the medians are close to the means, displaying diffusive spatial profiles. At strong monitoring, we find that the median density and current develop a domain-wall and single-peak profile, respectively, which are suggestive of a Zeno-like localization in typical quantum trajectories. While we are not able to identify a sharp phase transition as a function of the monitoring rate, our work highlights the usefulness of characterizing typical behavior beyond the averaged values in the context of monitored many-body quantum dynamics.
\end{abstract}

\maketitle

\section{Introduction}
\label{sec:intro}

An effective way of studying transport in many-body quantum systems is to induce a nonequilibrium steady state in the bulk by means of an external boundary driving. In one dimensional (1D) systems, the problem can be described by a Lindblad mater equation~\cite{breuer2002theory}, where nonzero currents are induced by asymmetric boundary driving terms~\cite{bertini2021finitetemperature}. This simple setting allows one to investigate a rich phenomenology, including anomalous transport and ergodicity-breaking effects arising \emph{e.g.} due to integrability~\cite{bertini2021finitetemperature, alba2021generalized}, kinetic constraints~\cite{sala2020ergodicity,feldmeier2020anomalous,singh2021subdiffusionandmanybody}, or disorder~\cite{abanin2019colloquium}.

In addition to boundary driving, one can also consider a bulk coupling to the environment. Simple Lindblad equations corresponding to this setting have been widely studied in the past two decades, especially in $1$D~\cite{znidaric2013transportina,jin2022exactdescription,jin2022semiclassicaltheoryof,jin2020generictransportformula,prosen2012diffusivehightemperaturetransport,prosen2009matrixproductsimulations,znidaric2019nonequilibriumsteadystatekubo}. These works have established that even weak decoherence-inducing terms typically lead to diffusive transport, a generic feature of both classical and quantum systems~\cite{bertini2015macroscopic,znidaric2020entanglementgrowthindiffusive,richter2022transportandentanglement,znidaric2020weakintegrabilitybreaking}. 

In this work we consider a different but closely related problem, where the bulk of a boundary-driven $1$D system is continuously monitored via weak-measurement processes~\cite{Wiseman2009,jacobs2006a,brun2002a}. While the averaged dynamics can be described by a standard Lindblad equation, leading to diffusive average transport, distinct histories of measurement outcomes define an ensemble of \emph{quantum trajectories}, displaying a much more interesting behavior. Our work is motivated by the recent literature on entanglement measurement-induced phase transitions (MIPTs)~\cite{skinner2019measurementinducedphase,li2018quantumzenoeffect,li2019measurementdrivenentanglement,noel2022measurement,koh2023measurement}, providing striking examples of how individual quantum trajectories may display  new phenomenology beyond the standard Lindbladian framework~\cite{fisher2022randomquantumcircuits,potter2022entanglementdynamicsin}. However, in contrast to most of the work in this literature which has investigated entanglement-related and quantum information aspects, we will exclusively focus on transport.

\begin{figure}[t!]
	\centering
	\includegraphics[width=\columnwidth]{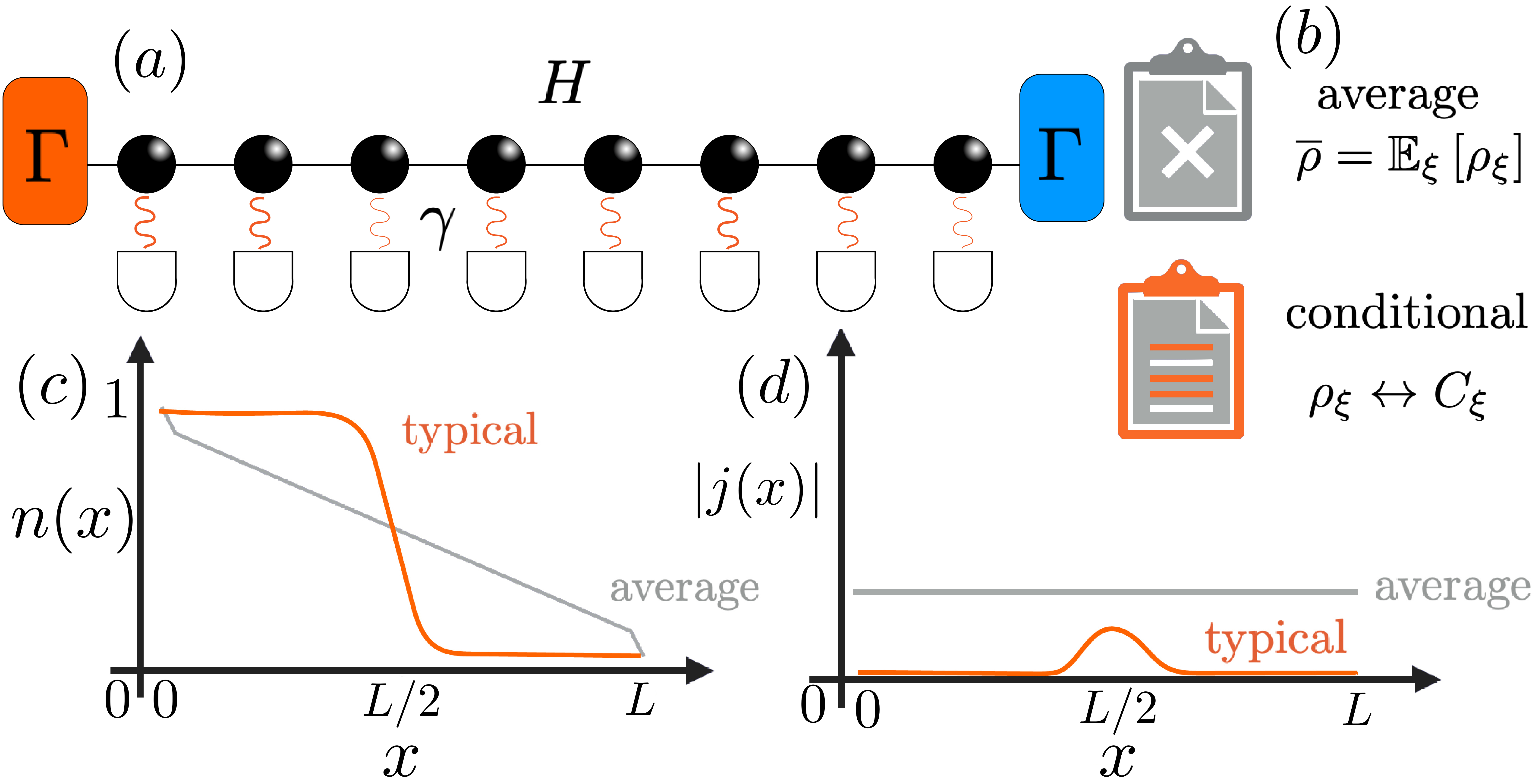}
	\caption{ Cartoon of the setup. (a) A fermionic chain is subject to the Hamiltonian $H$, left/right boundary driving and continuous bulk monitoring, with strength $\Gamma$ and $\gamma$, respectively. (b) The average state $\overline{\rho}$ is obtained by neglecting the measurement register and is described by the so-called dephasing model~\cite{turkeshi2021diffusionandthermalization}. 
		We compare the average density matrix with the conditional one $\rho_\xi$, depending on the history of measurement outcomes $\{\xi_{m,t}\}$. (c) At strong monitoring, the mean density displays a linear profile, whereas the locally typical density (associated with the median) is characterized by a domain-wall profile.
		(d) The average current is constant, while the locally typical current is inhomogeneous, displaying a single-peak profile. \label{fig:scheme}}
\end{figure}

We study a prototypical model of non-interacting free fermions featuring both boundary driving (in the form of particle injection-depletion) and bulk-monitoring of the particle density. This model was first considered in Ref.~\cite{bernard2018transport} and very recently in Ref.~\cite{turkeshi2022enhancedentanglementnegativity}, studying how the entanglement negativity is affected by the monitoring strength.  We also note that, in the absence of boundary driving terms, the model coincides with that studied in Refs.~\cite{cao2019entanglement,alberton2021entanglement,coppola2022growth,buchhold2021effective, poboiko2023theory}, in the context of MIPTs. 

It is important to stress that monitoring makes the dynamics highly non-trivial and difficult to study, even in the absence of interactions. For instance, it has been shown that the entanglement dynamics of monitored free fermions display a subtle behavior, with transitions and crossovers between phases with sub-extensive scaling~
\cite{cao2019entanglement,nahum2020entanglementanddynamics,tang2021quantumcriticalityin,van2021entanglemententropyscaling,lang2020entanglementtransitionin,fidkowski2021howdynamicalquantum,coppola2022growth,loio2023purification,boorman2022diagonisticsofentanglement,botzung2021engineereddissipationinduced,buchhold2021effective,turkeshi2021measurementinduced,turkeshi2022entanglementtransitionsfrom,turkeshi2022entanglementandcorrelation,gal2022volumetoarea,fleckenstein2022nonhermitiantopology,kells2021topologicaltransitionswith,piccitto2022entanglementtransitionsin,kawabata2022entanglementphasetransition,zhang2021emergentreplica,zhang2022universalentanglementtransitions,muller2022measurementinduceddark}. This phenomenon was quantitatively understood only very recently, exploiting a mapping to effective non-linear-sigma-model field theories~\cite{jian2023criticality,jian2023measurementinduced,fava2023nonlinear,poboiko2023theory}, see also~\cite{buchhold2021effective,yang2023keldysh}. 

In the presence of boundary driving, one may ask how the properties of the particle density and current along individual quantum trajectories are altered by the measurement rates. 
This question is particularly natural in light of the apparent competition between different effects: on the one hand, the driving forces a particle flow through the system; on the other hand, strong monitoring tends to pin the state toward an eigenstate of the measurement operation (pointer states). Therefore, it is natural to expect qualitative differences in the statistics of transport properties, similar to what happens for the entanglement negativity~\cite{turkeshi2022enhancedentanglementnegativity}.

In this work, we study the full probability distribution functions of the local particle density and current, focusing on the non-equilibrium  stationary regime of the model. Our main result is to highlight a qualitative difference between the average and typical transport. The former is determined the average density matrix, which is governed by a Lindblad master equation displaying diffusive features~\cite{esposito2005emergence,esposito2005exactly,eisler2011crossover,znidaric2010dephasinginduceddiffusivetransport,znidaric2010exactsolution,znidaric2011transportpropertiesofa,turkeshi2021diffusionandthermalization,turkeshi2022destructionoflocalization,scott2021subdiffusionina,znidaric2014exact,znidaric2014large}. The latter, which we identify with the statistical median, is computed out of the full probability distribution functions  in the ensemble of quantum trajectories, going beyond the Lindbladian description. We find that, at weak monitoring, the medians of local density and current are close to the means, displaying diffusive spatial profiles. At strong monitoring, we find that the median density and current develop a domain-wall and single-peak profile, respectively, which are suggestive of a Zeno-like localization in typical quantum trajectories. While we are not able to identify a sharp phase transition as a function of the monitoring rate, our work highlights the usefulness of characterizing typical behavior beyond the averaged values in the context of monitored many-body quantum dynamics.

The rest of this paper is organized as follows. In Sec.~\ref{sec:fermions},  we introduce the model and the protocol we consider. In Sec.~\ref{sec:observables} we introduce the observables of interest, including the local density, current and their statisics.
We present our numerical results in Sec.~\ref{sec:numerics}, specifically on the statistics of particle density and current (Sec.~\ref{sec:stat_density_current}) and their average versus typical behavior.  Our conclusions are consigned to Sec.~\ref{sec:conclusion}.

\section{Model, methods, and observables} \label{sec:fermions}

We study a 1D system of non-interacting fermions, where particles are injected and depleted at the left and right boundaries, respectively. We consider monitoring the local particle density, in the limit of weak but frequent measurements, as realized in, \textit{e.g.}, homodyne detection or the quantum-state diffusion protocol~\cite{gisin1992quantum,plenio1998the}, cf.~Fig.~\ref{fig:scheme}. Due to the boundary terms, the system is open and described by a density matrix. Its evolution is stochastic, due to the monitoring process. Thus, the overall evolution is encoded by the following stochastic master equation (SME)
\begin{align}
        d\hat{\rho}_\xi &= dt \mathcal{L}[\hat{\rho}_\xi] + \sum_{m=1}^L \left(d\mathcal{J}_m[\hat{\rho}_\xi]+\mathcal{D}_{m}[\hat{\rho}_\xi]\right)\,,
        \label{eq:sse}\\
         \mathcal{L}[\circ] &= -i [\hat{H},\circ] + \mathcal{D}_\mathrm{bnd}[\circ]\,.\label{eq:sse2}
\end{align}

\noindent Eq.~\eqref{eq:sse2} gives the first term in Eq.~\eqref{eq:sse} and describes the deterministic evolution of the system. Its first term is the coherent dynamics driven by the Hamiltonian
\begin{equation}
    \hat{H} = -\sum_{m=1}^{L-1} (\hat{c}_m^\dagger \hat{c}_{m+1} + \hat{c}_{m+1}^\dagger \hat{c}_m)\,,\label{eq:ham}
\end{equation}
where $\hat{c}_m$ and $\hat{c}_n^\dagger$ are the canonical fermionic operators satisfying $\{ \hat{c}_m,\hat{c}_n\} = 0$, $\{ \hat{c}_m,\hat{c}^\dagger_n\}=\delta_{m,n}$. The second term in Eq.~\eqref{eq:sse2} is the boundary Lindblad dissipator describing injection and depletion of particles,
\begin{align}
    \mathcal{D}_\mathrm{bnd}[\circ] &= \Gamma\left(2\hat{c}^\dagger_1 \circ \hat{c}_1 - \{\hat{c}_1 \hat{c}^\dagger_1 ,\circ\}\right)\nonumber \\
    &\qquad {+\Gamma\left( 2\hat{c}_L \circ \hat{c}^\dagger_L -\{ \hat{c}^\dagger_L \hat{c}_L ,\circ\}\right).}
\end{align}
The continuous monitoring of the local density $\hat{n}_i= \hat{c}^\dagger_i \hat{c}_{i}$ has two effects on the evolution of the mixed state of the system, as seen in the last term in 
Eq.~(\ref{eq:sse}). It introduces a local stochastic term $d\mathcal{J}_m[\circ]$,
\begin{equation}
d\mathcal{J}_m[\circ] \equiv \sqrt{\gamma} d\xi_t^m \{ \hat{n}_m-\langle \hat{n}_m\rangle_\xi,\circ\}\,,\label{eq:backfeed}
\end{equation}
and also gives rise to a dephasing term originating from the environment backaction 
\begin{equation}
\mathcal{D}_{m}[\circ] = -\frac{\gamma}{2} [\hat{n}_m,[\hat{n}_m,\circ]]\,.\label{eq:bulkdep}
\end{equation}
Here, we defined the expectation value $\langle \circ\rangle_\xi = \mathrm{tr}(\hat{\rho}_\xi \circ)$ and the stochastic real variable $d\xi^m_t$, satisfying the standard rules from \^Ito calculus $\overline{d\xi^m_t} = 0 $ and $d\xi^m_t d\xi^l_{t} = dt \delta_{m,l}$, see \emph{e.g.}~\cite{bauer2017stochastic} and references therein. $d\xi^m_t$ can be interpreted as an infinitesimal fluctuating noise term. We emphasize that bulk diffusion and stochastic terms are controlled by the same scale $\gamma$ and should not be understood as two independent processes competing as in Ref.~\cite{ladewig2022monitoredopenfermion}. In this work, we follow the standard terminology and define the set $\{\rho_{\xi}(t)\}$ as the ensemble of quantum trajectories, each labeled by a specific history of measurement outcomes $\{\xi_{m,t}\}$.

Due to the properties of the  \^Ito noise which is uncorrelated at different times, the average of Eq.~\eqref{eq:sse} yields the following Lindblad equation
\begin{equation}
    \frac{d}{dt}\hat{\overline{\rho}} = -i [H,\hat{\overline{\rho}}] +
    \mathcal{D}_\mathrm{bnd}[\hat{\overline{\rho}}]+\mathcal{D}_\mathrm{bulk}[\hat{\overline{\rho}}]\,,
    \label{eq:lindblad}
\end{equation}
with $\mathcal{D}_\mathrm{bulk}[\circ]=\sum_{m=1}^L\mathcal{D}_{m}[\circ]$, and $\mathcal{D}_{m}[\circ]$ defined in Eq.~(\ref{eq:bulkdep}). As mention, this Lindblad equation is well studied. In fact, it is analytically tractable: the $k$-point correlation functions constitute a hierarchy of decoupled equations of motion, each depending only on the set of $r$-point correlation functions with $r\leq k$~\cite{esposito2005emergence,esposito2005exactly,eisler2011crossover,turkeshi2021diffusionandthermalization}. In this work, however, we will be interested in the full ensemble of quantum trajectories, requiring us to go beyond the Lindbladian framework.

\subsection{Continuity equation for monitored dynamics}\label{sec:non_continuity}

In the absence of monitoring, the bulk Hamiltonian commutes with the total number of particles $\hat{N}=\sum_m \hat{n}_m$. From the continuity equation characterizing the Hamiltonian dynamics 
\begin{equation}\label{eq:continuity_eq}
\partial_t \hat{n}_x=-i[\hat{n}_x,\hat{H}] = -(\hat{j}_{x+1}-\hat{j}_x)\,,
\end{equation}
we obtain the well-known expression for the current density
\begin{equation}
{ \hat{j}_m\equiv i (\hat{c}_{m-1}^\dagger \hat{c}_m - \hat{c}^\dagger_m \hat{c}_{m-1})}\,.
\end{equation} 
Eq.~\eqref{eq:continuity_eq} implies that the expectation value of the local current in the stationary regime must be constant along the chain~\cite{znidaric2010exactsolution}. 

Monitoring significantly modifies this picture. Most prominently, the continuity equation~\eqref{eq:continuity_eq} has to be modified. To see this, it is useful to write down the derivative of the local density in the bulk of the system, as generated by the SME~\eqref{eq:sse}. Using the fact that the dynamics is Gaussian~\cite{bravyi2005lagrangian}, cf. Sec.~\ref{sec:numerics}, we can derive
\begin{align}
     dn_m(\xi) &= -[j_m(\xi) - j_{m-1}(\xi)]dt + dJ_m,\label{eq:dn_m}\\
     dJ_m &= 2 d\xi_m n_m(\xi) - 2\sum_k C_{m,k}(\xi) d\xi_k C_{k,m}(\xi)\label{eq:zio}\,,
\end{align}
where we used the notation $n_m(\xi)=\langle \hat{n}_m\rangle_\xi$, $j_m(\xi)  = \langle \hat{j}_m\rangle_\xi $ and $C_{k,m}(\xi)=\langle c^\dagger_k c_m\rangle_{\xi}$.  Eq.~\eqref{eq:dn_m} takes the form of a generalized continuity equation where a change in time in the local density corresponds to a current flowing to the neighboring site and to a stochastic fluctuating term $dJ_m$.
Taking the average over the noise, we get $\overline{dJ_m} = 0$ and obtain an exact continuity equation for the average current. However the fluctuating current $dJ_m$ at site $m$ is not zero along individual quantum trajectories. 

The term $dJ_m$ comes entirely from the non-unitary noise term in the SME. We can interpret $dJ_m$ as a fluctuating charge flowing out of (into) the measurement apparatus and entering (leaving) the system. This observation will be useful to interpret our results on transport statistics in the next section. In order to avoid potential confusion, however, we emphasise that this just a mathematical interpretation of the formula, but one should not claim that $dJ_m$ is associated to a measurable, physical flow of charge in or out of the system/detector.

While the fluctuating local current $j_m$ does not satisfy a conventional continuity equation, it still provides some information on how the charge is redistributed coherently across the lattice. For this reason in the next sections we will present results on the statistics of the trajectory-resolved local density and local current, $n_m(\xi)$ and $j_m(\xi)$.

\subsection{Average behavior and typical trajectories}
\label{sec:observables}

The average over quantum trajectories of any linear functional $\mathcal{L}$ of the system density matrix can be rewritten in terms of the average state $\hat{\overline{\rho}}\equiv \mathbb{E}_\xi [\hat{\rho}_\xi]$, \emph{i.e.} $\overline{\mathcal{L}[\hat{\rho}_\xi]}=\mathcal{L}[\hat{\overline{\rho}}]$. Conversely, for a non-linear functional $\mathcal{F}$, we have in general $\overline{\mathcal{F}[\hat{\rho}_\xi]}\neq \mathcal{F}[\hat{\overline{\rho}}]$. Setting $A(\xi)=\langle\hat{A}\rangle_\xi$, with $\hat{A} =\hat{j}_m$, $\hat{n}_m$, we are interested in the probability distribution function~\cite{presilla1996measurement,biella2021manybodyquantumzeno,tirrito2022fullcountingstatistics}
\begin{equation}\label{eq:ppaper}
	P_t(A;a) = \mathbb{E}_\xi  \left[\delta(\langle \hat{A}\rangle_\xi-a)\right]\,,
\end{equation}
which is as a highly non-linear functional of $\rho_\xi$ and, therefore, beyond the reach of the Lindbladian formalism.

The probability $P_t(A;a)$ is fully characterized by its statistical moments $K_{n}(A)$, which can be obtained in terms of the derivatives of the generating function
\begin{equation}
	G_t(A;\lambda) = \int da e^{-\lambda a} P_t(\hat{A};a)\,.
\end{equation}
Contrary to the full-counting statistics, typically studied in many-body physics, which quantifies the fluctuations of the measurement outcome of an observable in a given state (see Ref.~\cite{joaocosta:Thesis:2022}), $G_t(A;\lambda)$ characterizes the statistical fluctuations of the expectation values $A(\xi)$ in the ensemble of quantum trajectories $\{\rho_\xi\}$. For instance, the variance is given by
\begin{align}
	(\Delta A)^2:&=K_2(A)-K_1(A)^2\nonumber\\
	&= \mathbb{E}_\xi[\langle\hat{A}\rangle_\xi^2]- \mathbb{E}_\xi[\langle\hat{A}\rangle_\xi]^2\,.
\end{align}

In the next section, we will study the properties of $P_t(A;a)$. We will be interested in the typical value of $A(\xi)$ in the ensemble of the quantum trajectories. While the notion of the typical value of a probability distribution function is not uniquely defined (and sometimes taken as a synonym of the average value), in this work we associate it to the median of $P_t(A;a)$, which is defined as the value $\langle \langle A\rangle \rangle$ such that 
\begin{equation}
	\int_{-\infty}^{\langle \langle A\rangle \rangle} P_t(A;a) da = 1/2\,.
\end{equation}
The median is more naturally interpreted as an indicator of typical behavior than the mean when the probability distribution function is highly skewed. 
We will show that this is the case for the local particle density and current in the limit of strong monitoring. 

\subsection{Evolution of the covariance matrix}
\label{subsec:evocorrmat}

In this work, we perform exact numerical computations based on the formalism of fermionic Gaussian states~\cite{bravyi2005lagrangian}, following the approach used in Ref.~\cite{turkeshi2022enhancedentanglementnegativity}. We briefly review it here for completeness.

Fermionic Gaussian states are defined by the fact that they satisfy Wick's theorem~\cite{bravyi2005lagrangian}, and all correlation functions are uniquely determined by the so-called covariance matrix
\begin{equation}\label{eq:covmat}
	C_{m,l}= {\rm tr}\left[ \rho \hat{c}_m^\dagger \hat{c}_l\right]\,. 
\end{equation}
Crucially, the SME~\eqref{eq:sse} preserves Gaussianity~\cite{turkeshi2022enhancedentanglementnegativity}. Namely, an initial Gaussian state remains Gaussian along each of the quantum trajectories. Therefore, one can derive an efficient description of the dynamics in terms of the system covariance matrix. Using \^Ito calculus and Wick's theorem, one finds~\cite{turkeshi2022enhancedentanglementnegativity}
\begin{widetext}
\begin{align}
    dC_{m,l} &= i dt \left(C_{m-1,l} + C_{m+1,l} - C_{m,l+1} - C_{m,l-1}\right) - \gamma (1-\delta_{m,l})C_{m,l} dt + \sqrt{\gamma}(d\xi^m_t +d\xi^l_t) C_{m,l} \nonumber\\
    -& 2\sqrt{\gamma}\sum_{r=1}^L C_{m,r} d\xi^r_t C_{r,l}+ 2 \Gamma_L \delta_{m,1}\delta_{l,1} dt - \Gamma (\delta_{m,1} + \delta_{l,1} + \delta_{l,L} + \delta_{m,L}) C_{m,l} dt\,.
\label{eq:ssecmat}
\end{align}
\end{widetext}
Following~\cite{turkeshi2022enhancedentanglementnegativity}, we solve Eq.~\eqref{eq:ssecmat} by discretizing time, and splitting each time step into two subsequent intermediate steps (one can check that the discretization error in this procedure is vanishing in the limit where the discrete time interval approaches zero). In the first step, we evolve the system by the Hamiltonian $H$ and the boundary driving. This is done by integrating over a time interval $\Delta t$ (for instance using the Runge-Kutta method) the differential equation
\begin{equation}
    \frac{d}{dt} C(t) = \mathbb{L}[C(t)] + \mathbb{P},\label{eq:nonoise}
\end{equation}
where the linear opreator $\mathbb{L}$ and the matrix $\mathbb{P}$ are obtained from Eq.~\eqref{eq:ssecmat} setting $\gamma,d\xi^m_t = 0$. We denote the covariance matrix obtained in this way by $\tilde{C}(t+\Delta t)$. As a second step, we implement the non-unitary dynamics corresponding to the bulk measurement process. For each site, this is given by the transformation [cf.~Eq.~\eqref{eq:sse}]
\begin{subequations}
\label{eq:transrho}
\begin{align}
    \hat{\rho}' &=  \frac{\hat{T}_m\hat{\rho}\hat{T}_m}{\mathrm{tr}(\hat{T}^2_m\hat{\rho})},\\
    \hat{T}_m&\equiv \exp \left[ \sqrt{\gamma}d\xi^m_t \hat{n}_m -\gamma \Delta t (\hat{n}_m - \langle \hat{n}_m\rangle_\xi )^2\right].
\end{align}
\end{subequations}
In the numerical implementation, the random variables $\xi^{m}_t$ are independently drawn from a Gaussian distribution with zero average and variance ${ \mathbb{E}[(\xi_t^{m})^2]=\Delta t}$. Eq.~\eqref{eq:transrho} can be rewritten uniquely in terms of the covariance matrix, using the theory of fermionic Gaussian states~\cite{bravyi2005lagrangian,fidkowski2021howdynamicalquantum}.  After a lengthy but straightforward computation, we can write the transformed covariance matrix as $C(t+\Delta t) = \mathbb{J}_1\circ\dots \mathbb{J}_L [\tilde{C}(t+\Delta t)]$, with
\begin{align}
    \mathbb{J}_m(C)& =\! P^{(m)}\!\left[ C + x_m\!\left(E^{(m)} C + C E^{(m)} 
    \!-\! 2 C E^{(m)}  C\right)\right.\nonumber\\
    -&\left.\frac{x_m+1}{2} E^{(m)} \right] P^{(m)} + \frac{\tanh(\epsilon_m)+1}{2} E^{(m)},
\end{align}
where we defined the following matrix and coefficients
\begin{align}
    [D^{(m)}]_{l,r} &= \delta_{l,r} \left(\frac{1}{\cosh(\epsilon_j)} \delta_{m,l} + (1-\delta_{m,l}\right)\,,\\
    [E^{(m)}]_{l,r}  &= \delta_{m,l} \delta_{m,r}, \\
    \epsilon_m &= d\xi^m_t + (2C_{m,m}-1) \Delta t\,,\\ 
    x_m &= \frac{\tanh{(\epsilon_m)}}{1-(1-C_{m,m})\tanh(\epsilon_m)}\,.
\end{align}
By repeating this procedure for each time step, we obtain the evolved covariance matrix $C(t)$ for one particular history of measurement outcomes $\{\xi_{m,t}\}$. The full probability distribution function is obtained by sampling different quantum trajectories, repeating the full time evolution many times (each time new random numbers are generated). The computational cost of this method scales only polynomially in $L$ and $t$ for a single quantum trajectory, allowing us to reach large system sizes and simulation times. 

\section{Numerical results}\label{sec:numerics}

In this section we present the results of our numerical simulations. For convenience, we initialize the system in the infinite-temperature state, which is Gaussian and described by the covariance matrix $C = \openone_{L}/2$. The choice of the initial state is immaterial, as long as we are only interested in the late-time stationary regime, which is reached at times $t\sim (L/\gamma)^2$~\cite{turkeshi2022enhancedentanglementnegativity}. Note that, in this regime, the probability distribution function in Eq.~\eqref{eq:ppaper} becomes independent of time.  For our numerical simulations, we chose $\Delta t=0.05$, and verified that our results are stable upon decreasing $\Delta t $ further. For each value of the system parameters $\gamma$, $\Gamma$, and $L$, we have sampled $\mathcal{N}_\mathrm{traj} = 10^5$ quantum trajectories. 

\subsection{Trajectory-resolved vs averaged values}

We begin by discussing the qualitative features of the profiles of  $j_m(\xi)$ and $n_m(\xi)$ in individual quantum trajectories, as the measurement rate is increased. We focus on a sufficiently large time $t\gg (L/\gamma)^2$. To this end, it is useful to make a comparison with the averaged values in the so-called non-equilibirum steady state (NESS), which were analytically derived in Ref.~\cite{znidaric2010exactsolution}, reading
\begin{align}
    \lim_{t\to\infty} \overline{j_m(\xi)}& { =:j^{\rm NESS}=-\frac{1}{\Gamma+\Gamma^{-1} + (L-1)\gamma/2}}\nonumber\\
    &\simeq - \frac{2}{\gamma L},\label{eq:asympt_current}\\
    \lim_{t\to\infty} \overline{n_m(\xi)}&= 1 + j^\mathrm{NESS}\left[1+ (m-1)\frac{\gamma}{2} \right.\nonumber\\
    +&\left. \frac{1}{2}(\delta_{m,1}-\delta_{m,L})\right]\,.\label{eq:ave}
\end{align}
We see that the mean current and density profiles display a crossover from ballistic to diffusive transport as the system size increases. In the limit where $L$ is the largest length scale, the density linearly decreases from one end to the other, with a uniform current $j^\mathrm{NESS}\propto 1/L$. This is consistent with Fick's law, with the current approaching zero in the thermodynamic limit, as is typical for diffusive scaling~\cite{znidaric2010exactsolution,turkeshi2021diffusionandthermalization} [cf.~Fig.~\ref{fig:scheme}(c,d)]. 

\begin{figure}
	\centering
	\includegraphics[width=\columnwidth]{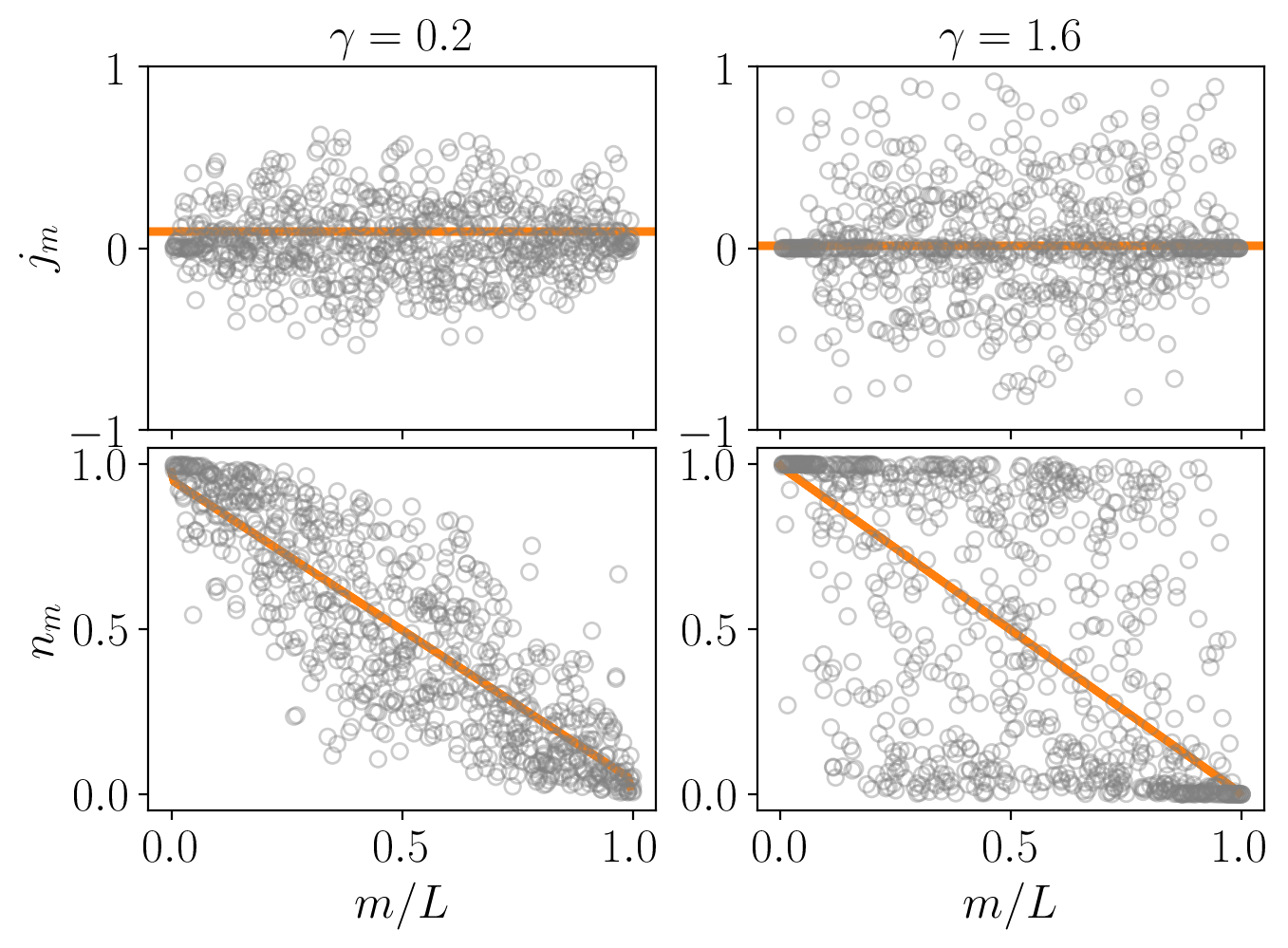}
	\caption{Comparison between the average values of $n_m$, $j_m$  (orange solid lines) and those in a randomly generated, individual quantum trajectory (grey markers) for various measurement rates and a sufficiently late time, $t\gg (L/\gamma)^2 $. Left panels: At low $\gamma$, the values in individual trajectories are typically close to the average ones. Right panels: For large $\gamma$, the mean values are strongly influenced by atypical configurations, and fluctuations are more pronounced. { Here we chose $\Gamma=1$, while the system size is $L=160$.}}
	\label{fig:realizations}
\end{figure}

In a single quantum trajectory, the profiles of $j_m(\xi)$ and $n_m(\xi)$ are not smooth due to the fluctuations induced by the measurements and we find that their qualitative features depend on the measurement rate. An example of our data is reported in Fig.~\ref{fig:realizations}. For weak measurement rate (left panels) we see that the amplitudes of fluctuations are small, and $j_m(\xi)$, $n_m(\xi)$ are close to their averaged values. Conversely, at large measurement rate (right panels), fluctuations are large. In particular we note that the probability distribution of $n_m$ has peaks near $0$ and $1$. The distribution of $j_m$ does not have peaks away from zero, but it displays tails that reach out towards $\pm 1$.

\begin{figure*}
    \centering
    \includegraphics[width=\textwidth]{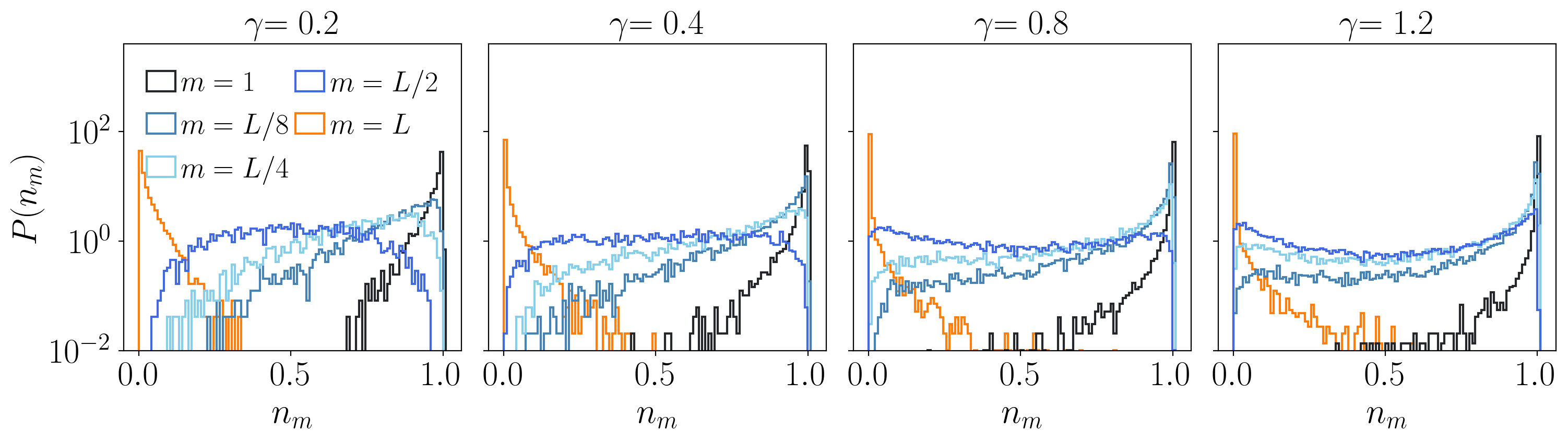}
    \includegraphics[width=\textwidth]{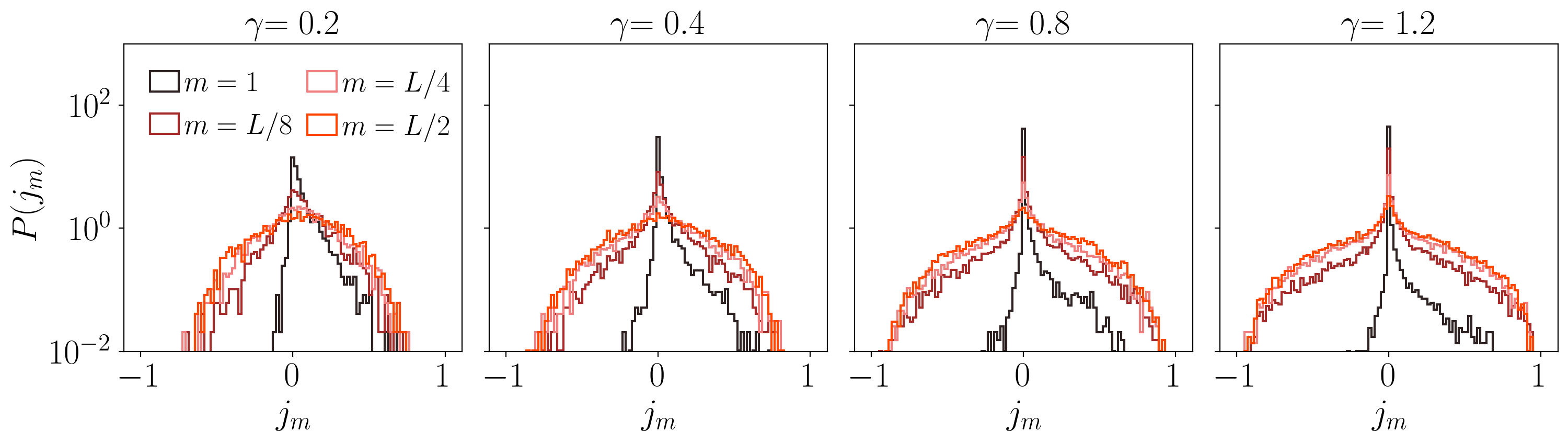}
    \caption{Histograms corresponding to the probability distribution functions of the particle density $P(n_m)$ (top) and current $P(j_m)$ (bottom). The plots are obtained by sampling the dynamics at sufficiently large times. We consider a chain of size $L=160$ with $\Gamma=1$ and vary the monitored strength $\gamma=0.2,0.4,0.8,1.2$. We compare the statistics at the boundaries $m=1,L$ and at different positions in the bulk, $m=L/8,L/4,L/2$. We have tested that the average density and current computed out of the data shown here match the analytic result for the Lindbladian steady values, up to the statistical accuracy.}  \label{fig:density}
\end{figure*}

\subsection{Statistics of Particle Density and Current}\label{sec:stat_density_current}
The previous discussion suggests that, depending on the measurement rate $\gamma$, the average properties of the state are dominated by either values close to the mean (small $\gamma$) or far from it (large $\gamma$). We make this observation quantitative by studying the full statistics of the observables of interest. To this end, we compute the probability distribution functions $P(j_m)$ and $P(n_m)$ [cf. Sec.~\ref{sec:observables}], which we plot in Fig.~\ref{fig:density} for different values of the monitoring rate $\gamma$ and position along the chain $m$, for a system of length $L=160$. 
From the plots, we can immediately appreciate qualitative differences between the boundaries and the bulk of the system.

Let us first focus on the low-$\gamma$ regime (left panels in Fig.~\ref{fig:density}). We see that the bulk current (computed at position $m=L/2$) is approximately normally distributed with a large variance. At the same time, the density has a nearly-flat histogram: in the limit of very low measurement rate, almost all particle configurations are equally likely, as the particles flow through the system. 
Conversely, the statistics of local particle density and current is very different at the edges of the chain. We see that the probability distribution functions of the particle density are increasingly skewed as $m$ is moved close to the boundaries, displaying exponentially suppressed tails. At the same time, $P(j_m)$ develops a sharp peak close to zero current and broad tails which are strongly asymmetric due to the presence of the pump/loss terms. 
As the measurement rate is increased (right panels in Fig.~\ref{fig:density}), the peak at the boundary becomes sharper, while the bulk develops fatter exponential tails and a more pronounced (albeit lower) zero-current peak. The distribution function $P(n_m)$ is also qualitatively different at large monitoring. In the bulk, the distribution develops two peaks at the values $n_m=0,1$, while the peaks at the boundaries are more pronounced. 

\subsection{Particle-density profiles}

\begin{figure}
    \centering
    \includegraphics[width=\columnwidth]{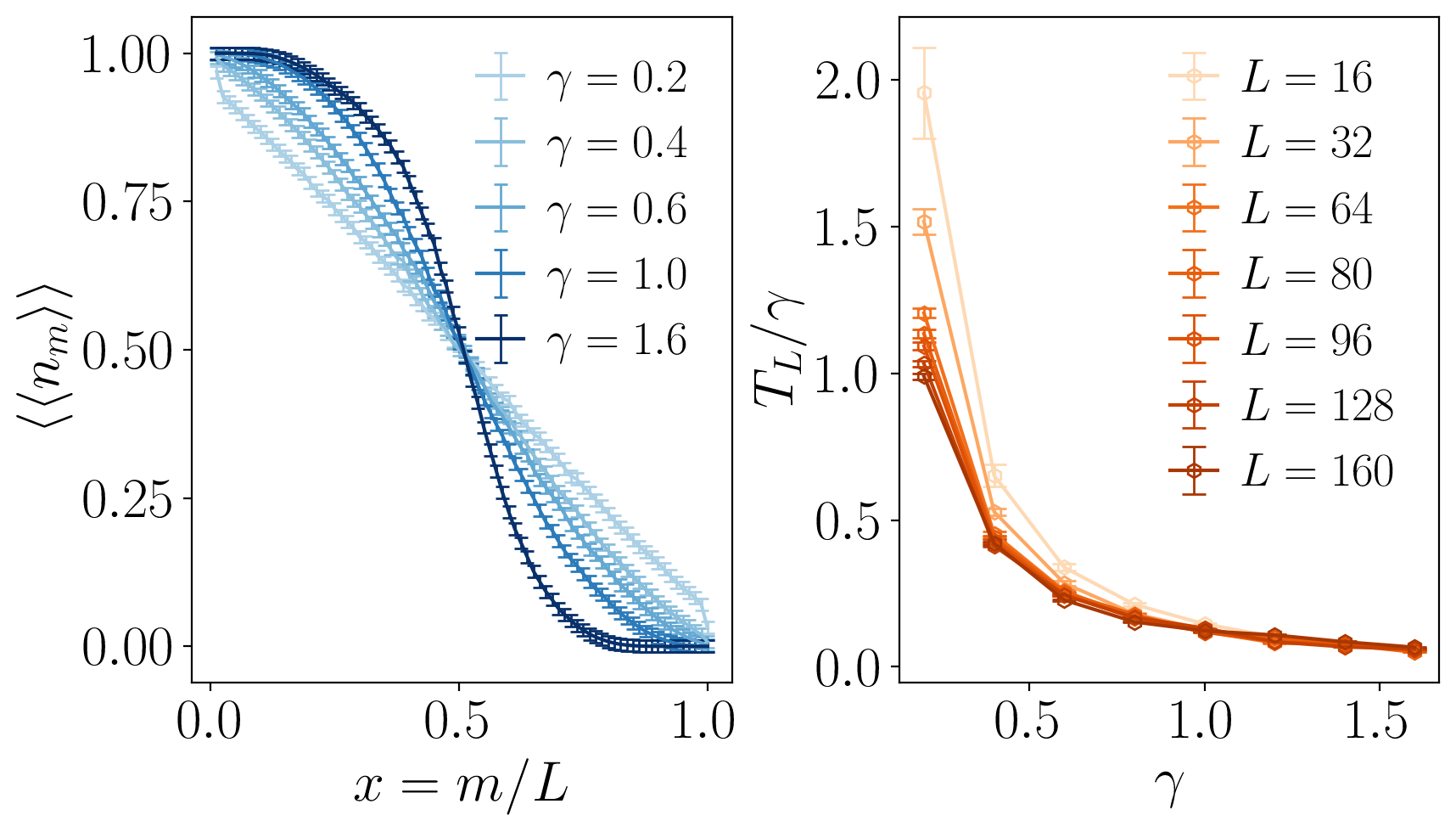}
    \caption{(Left) Typical density for various $\gamma$ (for $L=128$). We extract the finite size fit at each $L$ via  Eq.~\eqref{eq:fermifun}. (Right) Scaling of $T_L$ versus the measurement rate $\gamma$. Our data show a crossover between a localized behaviour with $\gamma\gtrsim 0.7$ and one with $\gamma\lesssim 0.7$ where the system is weakly delocalized.}
    \label{fig:nsub}
\end{figure}

We first extract from our data the typical behavior of the expectation values of the particle density, $n_m(\xi)$. As discussed in Sec.~\ref{sec:observables}, we do this by computing the median of its probability distribution function, which we denote by $\langle\langle n_m\rangle\rangle$.  Our results are reported in Fig.~\ref{fig:nsub}(a), showing the profiles of the locally typical density along the chain for different measurement rate values $\gamma$ and for fixed $L=128$. 

We see that the spatial profiles are qualitatively different from what is expected for conventional diffusive behavior~\cite{bertini2021finitetemperature}. While the density profiles is linear at small $\gamma$, being very close to the average one, it is deformed as $\gamma$ increases, developing a domain-wall profile as $\gamma\to\infty$. This is typical of localized behavior, with the particles being pinned near the edges to values corresponding to full and empty sites. We interpret this effect as a signature of Zeno-like localization due the interplay between monitoring process and boundary drive. The former tends to freeze the system in an eigenstate of the measurement operators with no charge fluctuations, while the latter fixes the density at the boundaries and breaks the symmetry between full and empty sites. On the other hand the average density profile at strong monitoring remains diffusive, with a slope independent of $\gamma$, see Eq.~\eqref{eq:ave}. 

For large but finite values of $\gamma$, the shape of the profiles is qualitatively similar to the one encountered in the presence of anomalous transport, interpolating between diffusive and localized behaviour~\cite{bertini2021finitetemperature}. In this case, the shapes of the profiles as a function of the scaling variable $x=m/L$ can be predicted in the large-$L$ limit, heuristically assuming  a space-dependent diffusion constant~\cite{znidarivc2011transport, znidaric2016diffusive, gullans2019localization}. Here we take a simpler empirical approach, and consider fitting the numerical data against the function 
\begin{equation}
    n(x) = \frac{1}{1+e^{\gamma (x-1/2)/T_\mathrm{eff}}}.
    \label{eq:fermifun}
\end{equation}
The fitting parameter $T_{\rm eff}$ should be interpreted as an effective temperature, competing with the tendency of the system to localize. As shown in Fig.~\ref{fig:nsub}(b), we find that, for large $\gamma$, Eq.~\eqref{eq:fermifun} captures well the behavior of the typical particle density, since the resulting fit becomes independent of the system sizes. At low and intermediate $\gamma$, $T_L = T_\mathrm{eff}(L)$ displays a finite-size scaling, slowly decreasing as $L$ increases. Therefore, our results suggest the existence of two different regimes as a function of $\gamma$ (depending on whether $T_{\rm eff}(L)$ depends or not on $L$). However, our data do not allow us to identify whether this change is associated to a true phase transition or crossover, as $L$ increases. For the available system sizes, the change in the behavior of $T_{\rm eff}$ appears to take place at $\gamma_c\sim 0.7$. Note that the smallest value of $\gamma$ chosen in Fig.~\ref{fig:nsub}(b) is dominated by finite-size effects for $L<128$. This can be seen from Eq.~\eqref{eq:asympt_current}, from which we see that the asymptotic diffusive behavior is realized for $L\gg 2/\gamma$. Similar finite-size effects should be expected in Fig.~\ref{fig:median_current_profiles}.

We further validate these results in App.~\ref{sec:app}, where the mode of $n_m$ is considered as another typicality indicator.

\subsection{Current-density profiles}

\begin{figure*}[t!]
    \centering
    \includegraphics[width=\textwidth]{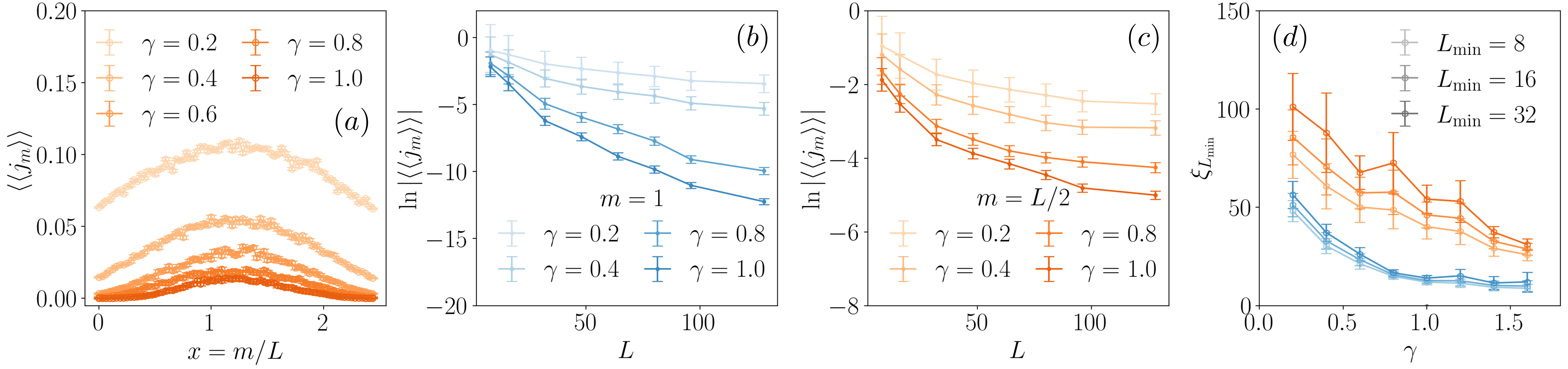}
    \caption{(a) Typical current for various measurement rates for $L=96$. (b)-(c) Log-log plots for $\langle\langle j_m\rangle\rangle$ for various $L$ and $\gamma$ at $m=1$ (b) and $m=L/2$ (c). We extract the estimate localization length via the fit in Eq.~\eqref{eq:fitj}, neglecting $L\le L_\mathrm{min}$ [where the values of $L_{\rm min}$ are reported in panel (d)]. (d) Scaling of the exponents with $\gamma$ at various $L_\mathrm{min}$ for $m=1$ (blue lines) and $m=L/2$ (orange lines).}
    \label{fig:median_current_profiles}
\end{figure*}

Next, we discuss the profiles of the typical current density $j_m(\xi)$, as quantified by the median of its probability distribution function (denoted by $\langle\langle j_m\rangle\rangle$). Our results are reported in Fig.~\ref{fig:median_current_profiles}(a). For small $\gamma$, $j_m$ is slowly varying in space, while upon increasing the rate it becomes strongly suppressed at the boundaries and non-zero only in the central region of the chain, deviating from the typical constant diffusive profile.
To be quantitative, we study the dependence of $\langle\langle j_m\rangle\rangle$ with the system-size $L$,  considering both the current close to the boundary [Fig.~\ref{fig:median_current_profiles}(b)] and in the bulk [Fig.~\ref{fig:median_current_profiles}(c)], for different measurement rates. In each case, we fit the numerical data against the function
\begin{equation}
    \langle\langle j_m\rangle\rangle \sim \exp(-L/\xi^\mathrm{loc}_m),\label{eq:fitj}
\end{equation}
Here $\xi^{\rm loc}_m$ can be interpreted as an effective space-dependent correlation length. In the limit of very large $\gamma$, we expect $\xi^\mathrm{loc}_m\to 0$, consistent with Zeno localization. The fits are performed neglecting system sizes $L\le L_\mathrm{min}$, for increasing $L_{\rm min}$, and considering only larger $L$ up to $L=160$. Examples of our numerical data for $\langle\langle j_m\rangle\rangle$ at the boundaries and in the bulk are reported in Figs.~\ref{fig:median_current_profiles}(b) and (c), while our results for $\xi_{\rm loc}$ are plotted in Fig.~\ref{fig:median_current_profiles}(d). We see that, at the sizes that we study, the decay of the current is not clearly exponential, the plots showing a non-zero curvature. Therefore, the fit in Eq.~\eqref{eq:fitj} should be taken with some care. Still, Fig.~\ref{fig:median_current_profiles}(d) shows that the fit is accurate in the limit of large $\gamma$.

The localization length appears to be smaller at the boundary, and in any case decreasing as $\gamma$ increases, as expected. At the boundary, we find $\xi^\mathrm{loc}\sim 1/\gamma$ for large gamma. Once again, we are not able to identify a sharp transition in the features of $\xi^\mathrm{loc}_m$ as function of $\gamma$. However, our results for the boundary correlation length suggest the existence of two regimes: for small $\gamma$, it appears to weakly increase with the system size, while for large $\gamma$ it is constant within the fitting error. Consistently with our findings for the particle density, the two regimes are separated by the value $\gamma\sim 0.7$, at least for the system sizes which we can reach. We also note that in the bulk no immediate difference emerges in the scaling of $\xi^{\rm loc}_m$.

It is interesting to interpret these results in light of the modified continuity equation for the charge discussed in Sec.~\ref{sec:non_continuity}. In fact, from a mathematical point of view, we see from Eq.~\eqref{eq:zio} that the amount of local particle density flowing randomly into the detector is controlled by the covariance matrix along that quantum trajectory. It is easy to see that the freezing of charge fluctuations at the boundaries implies a reduced net flow of charge entering or leaving the system, as we can see from the continuity equation. However, one should keep in mind that the non-unitarity of the monitored dynamics makes it difficult to interpret the transport behavior discussed here in terms of conventional hydrodynamics.

\section{Conclusions}\label{sec:conclusion}

In this work we have studied a prototypical model of a continuously monitored many-body system subject to external driving, and analyzed quantitatively its properties beyond the Lindbladian framework. Focusing on the late-time limit of standard transport observables, namely local particle density and current, we have shown how their probability distributions exhibit different qualitative features as a function of the monitoring rate. In particular, we have shown how the profiles of their medians are similar to those encountered in anomalous transport in stardard Lindbladian settings. In the limit of very large monitoring rate, we found that the profiles exhibit features which are typical of localized phases, with a domain-wall and a single-peak profile for the local particle density and current, respectively.  While we have not been able to identify a sharp phase transition as a function of the monitoring rate, our work highlights the usefulness of typicality probes beyond the mean value to extract useful information on the ensemble of quantum trajectories.

It would be interesting to substantiate our findings using recently-developed field-theoretical approaches for monitored free fermions, see \emph{e.g.} Refs.~\cite{poboiko2023theory,buchhold2021effective,fava2023nonlinear}. Another natural question is how transport features in individual quantum trajectories would be modified in the presence of additional unitary noise or interactions. We leave these questions for future works.

\begin{acknowledgements}
We acknowledge computational resources on the Colle\'ge de France IPH cluster. 
X.T. and M.S. were supported by the ANR grant ``NonEQuMat'' (ANR-19-CE47-0001).
\end{acknowledgements}

\begin{figure}[t!]
    \centering
    \includegraphics[width=\columnwidth]{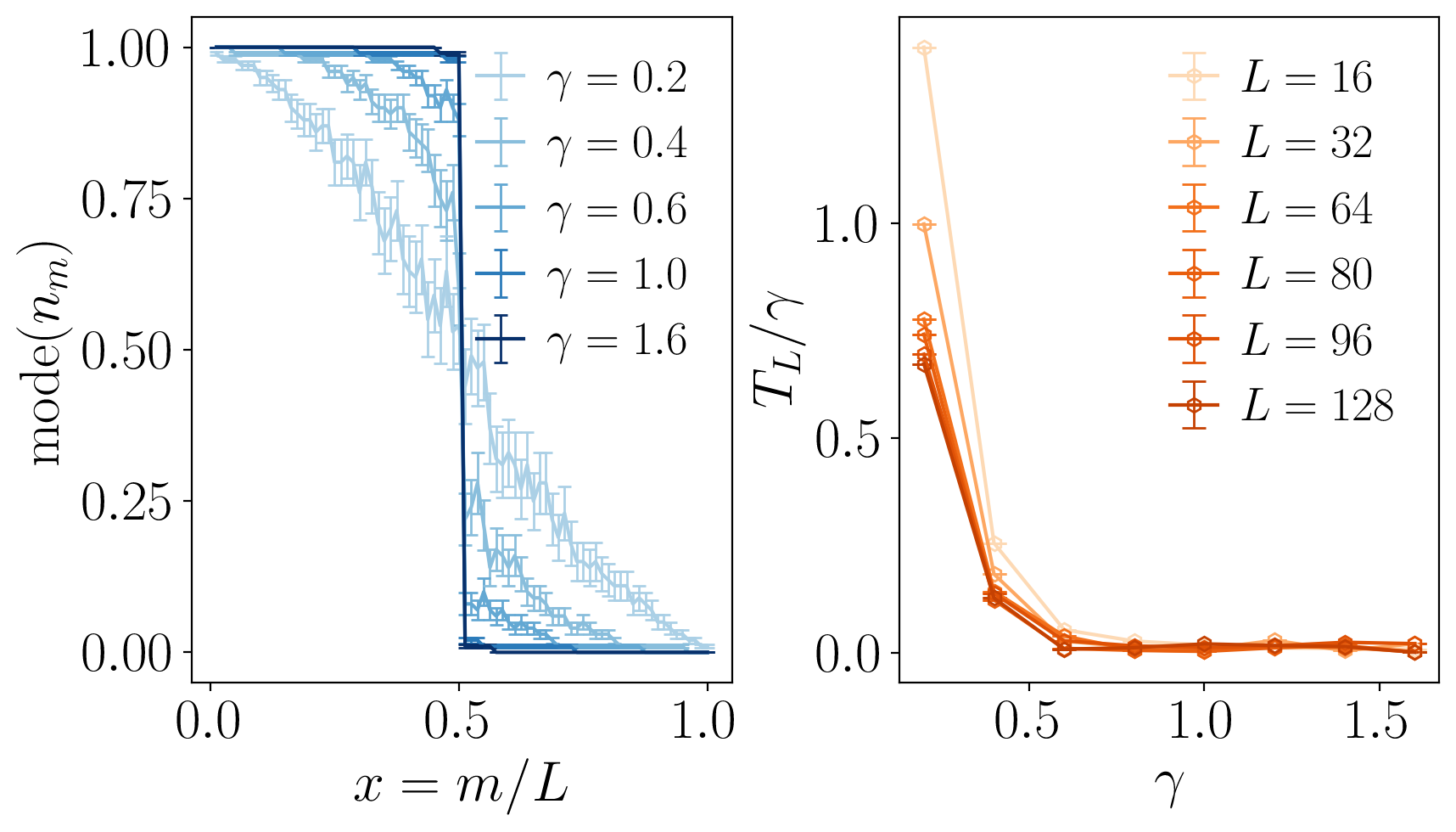}
    \caption{ (Left) Mode of $n_m$ for $L=128$ and various values of $\gamma$. The mode highlights a domain wall shape, consistent with the qualitative analysis discussed in the Main Text. (Right) Scaling of $T_L$  extracted from the mode versus the measurement rate $\gamma$. As for the median, our data show a crossover between a localized behavior and a delocalized one around $\gamma \sim 0.7$. }
    \label{fig:appendix}
\end{figure}

\appendix 
\section{Additional numerical results}
\label{sec:app}
In this section, we discuss the mode of $n_m$ as an additional indicator of typicality. This quantity, denoted as $\mathrm{mode}(n_m)$ is defined as the location of the maximum of the distribution $P(n_m)$.
The quantitative values of the mode depend drammatically on the data binning for the histograms, cf. Fig.~\ref{fig:density}. In this section, we consider two realizations of $n_m(\xi_1)$ and $n_m(\xi_2)$ the same if $|n_m(\xi_1)-n_m(\xi_2)|<0.3\times  10^{-2}\sim 1/\sqrt{\mathcal{N}_\mathrm{traj}}$. 

Our results, reported in Fig.~\ref{fig:appendix}, corroborate the conclusions in Sec.~\ref{sec:numerics}. The spatial profiles are qualitatively different from the conventional diffusive behavior, and at $\gamma\to\infty$ the density profiles reaches a pronounced domain wall behavior. 
We further validate the Zeno-like localization considering the fit in Eq.~\eqref{eq:fermifun} on the mode data. Our results are presented in Fig.~\ref{fig:appendix}(Right). Here $T_L\simeq 0$ for $\gamma\gtrsim 0.7$, and $T_L>0$ for $\gamma \lesssim 0.7$. 

Two comments are in order. First, the quantitative difference between median and mode is expected. The ensemble of quantum trajectory is not self-averaging, and different indicators highlight inequivalent aspects. 
Furthermore, quantitative conclusions on the mode should be taken with some care due to the aforementioned data binning-dependence. In contrast, average and median do not depend on the binning of the dataset, thus providing more robust statistical indicators.

\bibliography{entanglement.bib}
\bibliographystyle{apsrev4-2}

\end{document}